\documentclass[aps, prb, twocolumn, superscriptaddress, longbibliography]{revtex4-2}
\usepackage[colorlinks, linkcolor=blue, anchorcolor=blue, citecolor=blue]{hyperref}
\usepackage{amsmath}
\usepackage{graphicx}
\usepackage{braket}
\usepackage{multirow}
\usepackage{color}
\usepackage{soul}
\usepackage{diagbox}
\usepackage{rotating}

\usepackage{ulem}
\begin{document}

\title{Heavy-fermions in frustrated Hund's metal with portions of incipient flat-bands}
\author{Yilin Wang}\email{yilinwang@ustc.edu.cn}
\affiliation{School of Emerging Technology, University of Science and Technology of China, Hefei 230026, China}
\affiliation{Hefei National Laboratory, University of Science and Technology of China, Hefei 230088, China}
\affiliation{New Cornerstone Science Laboratory, University of Science and Technology of China, Hefei, 230026, China}

\date{\today}

\begin{abstract}
    Flat-bands induced by destructive interference of hoppings in frustrated lattices such as kagome metals, have been extensively studied in recent years. However, such flat-bands usually appear in small portions of Brillouin zone and are away from Fermi level in the non-interacting limit (dubbed as ``incipient"). Whether such incipient flat-band portions can induce very strong electron correlations is an open question. Recently, such incipient flat-bands are found in frustrated kagome CsCr$_3$Sb$_5$ and triangular CrB$_2$, which show moderately heavy-fermion behaviors and unconventional superconductivity. Here, by density functional theory plus dynamical mean-field theory calculations, we show that both compounds are typical Hund's metals. The incipient flat-band portions induce dips in hybridization functions, which further enhance the Kondo-like effect of Hund's metal, resulting in moderately heavy-fermion behaviors. This explains the origin of the heavy-fermion behaviors and pave the way for understanding the unconventional superconductivity in CsCr$_3$Sb$_5$ and CrB$_2$. Our work demonstrates a flexible route for generating $d$-electron heavy-fermion and for inducing unconventional superconductivity based on frustrated Hund's metals with portions of incipient flat-bands. A large family of materials with kagome, triangular, pyrochlore lattices consisting of Cr (d$^4$), Ru (d$^4$), Fe (d$^6$) ions are promising candidates. 
\end{abstract} 

\maketitle

\section{Introduction} 
In narrow/flat band systems, electron-electron interaction usually give rise to strongly correlated phenomena~\cite{Imada:1998,Laughlin:1983,Xiaogang:2011,Regnault:2011,MacDonald:2011,Caoyuan:2018}. For example, heavy-fermions are induced via Kondo effect in $f$-electron systems~\cite{Stewart:1984,Coleman:2007,Siqimiao:2010}. Flat-bands could be also induced in frustrated lattices by destructive interference of hoppings~\cite{syozi:1951,Mielke:1991,Sachdev:1992,Regnault:2022}, such as those extensively studied in kagome metals~\cite{Liu:2020,Kang:2020,Changgan:2022,Zhenyu:2018,Kang:2020ER,Yin:2019,Thomale:2013,Brian:2020,Checkelsky:2024}. However, in multiorbital $d$-electron materials, such flat-bands usually appear in small portions of Brillouin zone (BZ) and are away from Fermi level ($E_\text{F}$) (dubbed as incipient flat bands~\cite{Aoki:2020}). Whether such incipient flat-band portions can induce very strong electron correlations and even heavy-fermions is an open question. For example, although flat-bands are found near $E_{\text{F}}$ in kagome CoSn~\cite{Liu:2020}, the calculations yield a mass-enhancement less than 2~\cite{Liu:2020,Huangli:2022} and the transport experiment~\cite{Brian:2020} found a very small specific-heat coefficient $\gamma$ $\sim$3.68 mJ mol$^{-1}$ K$^{-2}$, indicating weak electron correlations.

In multiorbital $d$-electron systems, there is one class of strongly correlated metals dubbed as Hund's metals, where correlations mainly derive from Hund's coupling $J_{\text{H}}$, not Hubbard $U$~\cite{Werner:2008,haule:2009,medici:2011,yin:2011,YinZP:2011,georges:2013,Stadler:2015,stadler:2019,Kim:2022,Basov:2012,Medici:2017,Hirjibehedin:2015}. They usually appear when the electron filling is one-electron away from half-filling~\cite{medici:2011}. Canonical Hund's metals include iron-based superconductors (6 electrons/5 $d$-orbitals)~\cite{yin:2011,YinZP:2011,Hardy:2013,Shenzx:2013} and Sr$_2$RuO$_4$ (4 electrons/3 $t_{2g}$ orbitals)~\cite{Mravlje:2011,Dengxiaoyu:2016,Takeshi:2016,Karp:2020,Minjae:2018}. Signatures of Hund's metal are also found in nickelate superconductors~\cite{WYL:2020,Kangchangjong:2021,Sangkook:2021,Ouyang:2024,Jingxuan:2024}. Hund's metals show strong orbital differentiation~\cite{Medici:2009,Kugler:2019,Werner:2007,Yiming:2015,Kostin:2018,Nicola:2013,Zingl:2019,Koga:2005,Koga:2004,Kugler:2022},
 and it could be significantly amplified via larger $J_{\text{H}}$ even though the original orbital difference is not that large, which leads to coexistence of heavy and light electrons. Hund's coupling tends to form large local moments which leads to strong incoherence of electrons at high temperature. With decreasing temperature, Hund's metal undergoes a Kondo-like incoherence-to-coherence crossover by screening the local moments, and a renormalized Fermi-liquid with large quasi-particle mass emerges at very low temperature~\cite{haule:2009,Mravlje:2011,Hardy:2013,Stadler:2015,WYL:2020b,Coleman:2009,Drouin:2021,Drouin:2022}. In this sense, the Hund's metal physics is in the same spirit of Kondo effect. A question thus arises: whether the presence of incipient flat-band portions in a Hund's metal will further enhance the Kondo-like effect such that heavy-fermions are induced?

\begin{figure*}
    \centering
    \includegraphics[width=1.0\textwidth]{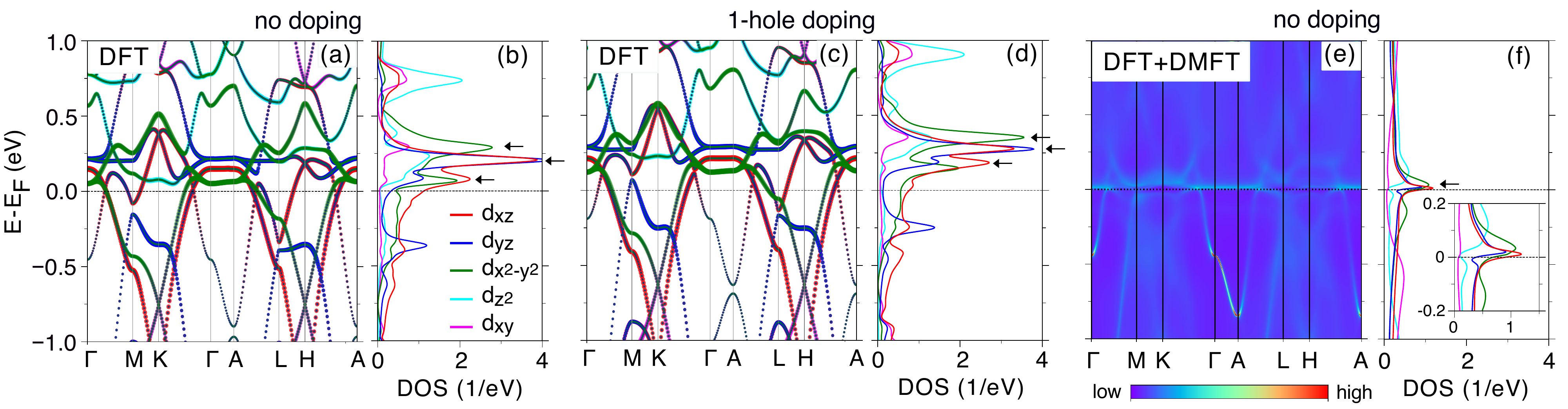}
    \caption{Incipient flat-bands in CsCr$_3$Sb$_5$. (a)-(d) DFT calculated band structures and density of states. (a),(b) Without doping. (c),(d) With one-hole doping. (e),(f) DFT+DMFT calculated spectra functions for the case without doping, at $U=5$ eV, $J_H=0.88$ eV and $T=100$ K. The black arrows in (b),(d),(f) indicate the positions of the incipient flat-bands with $d_{xz}$, $d_{yz}$ and $d_{x^2-y^2}$ characters, which are pushed further away from $E_{\text{F}}$ by one-hole doping. The inset in (f) shows enlarged DOS.
     }
    \label{fig:band}
\end{figure*}

In this work, we demonstrate the answer is yes, by performing density functional theory plus dynamical mean-field theory (DFT+DMFT)~\cite{Georges:1996,kotliar:2006,lichtenstein:2001} calculations on the recently discovered unconventional superconductors, CsCr$_3$Sb$_5$~\cite{Caoguanghan:2023} and CrB$_2$~\cite{Qiyanpeng:2021}, with frustrated kagome and triangular lattices, respectively. CsCr$_3$Sb$_5$ crystallizes into the CsV$_3$Sb$_5$-type structure~\cite{Ortiz:2019,Stephen:2020}. At ambient pressure, the resistivity of CsCr$_3$Sb$_5$ and CrB$_2$ show semiconducting~\cite{Caoguanghan:2023} and sublinear~\cite{Bauer:2014} behaviors, respectively, indicating bad-metals. Large specific-heat coefficient $\gamma$ $\sim$120 and 70 mJ mol$^{-1}$ K$^{-2}$ are found for CsCr$_3$Sb$_5$ and CrB$_2$, respectively, indicating possible moderately heavy-fermions~\cite{Caoguanghan:2023,Bauer:2014} in the normal states. CsCr$_3$Sb$_5$ undergoes a novel charge density wave (CDW) transition and then an anti-ferromagnetic (AFM) transition at low temperature ($\sim$54 K)~\cite{Caoguanghan:2023}. First-principle calculations suggest it is an altermagnet~\cite{Linhaiqing:2023}. CrB$_2$ undergoes an AFM transition at 88 K~\cite{Bauer:2014}. Under high pressure, the CDW and AFM orders are suppressed and superconductivity occur in both compounds~\cite{Caoguanghan:2023,Qiyanpeng:2021,IgorMazin:2023}. DFT calculations find portions of incipient flat-bands with Cr $d$-orbital characters in both compounds~\cite{Caoguanghan:2023,Linhaiqing:2023,IgorMazin:2023}.

Our calculations show that both compounds are typical Hund's metals with unusually large quasi-particle mass-enhancement $m^*/m^{\text{DFT}}$ (10 $\sim$ 20). We show that the incipient flat-band portions induce dips in hybridization functions, which further enhance the Kondo-like effect of Hund's metals, resulting in moderately heavy-fermion behaviors. 
This provides a clear picture for understanding the origin of the heavy-fermion behaviors in CsCr$_3$Sb$_5$ and CrB$_2$ and pave the way for understanding their unconventional superconductivity. Our work demonstrates a flexible route for generating $d$-electron heavy-fermions based on frustrated Hund's metals with portions of incipient flat-bands. A large family of materials with kagome, triangular, pyrochlore lattices consisting of Cr (d$^4$), Ru (d$^4$), Fe (d$^6$) ions are thus promising candidates.

\begin{figure*}
    \centering
    \includegraphics[width=1.0\textwidth]{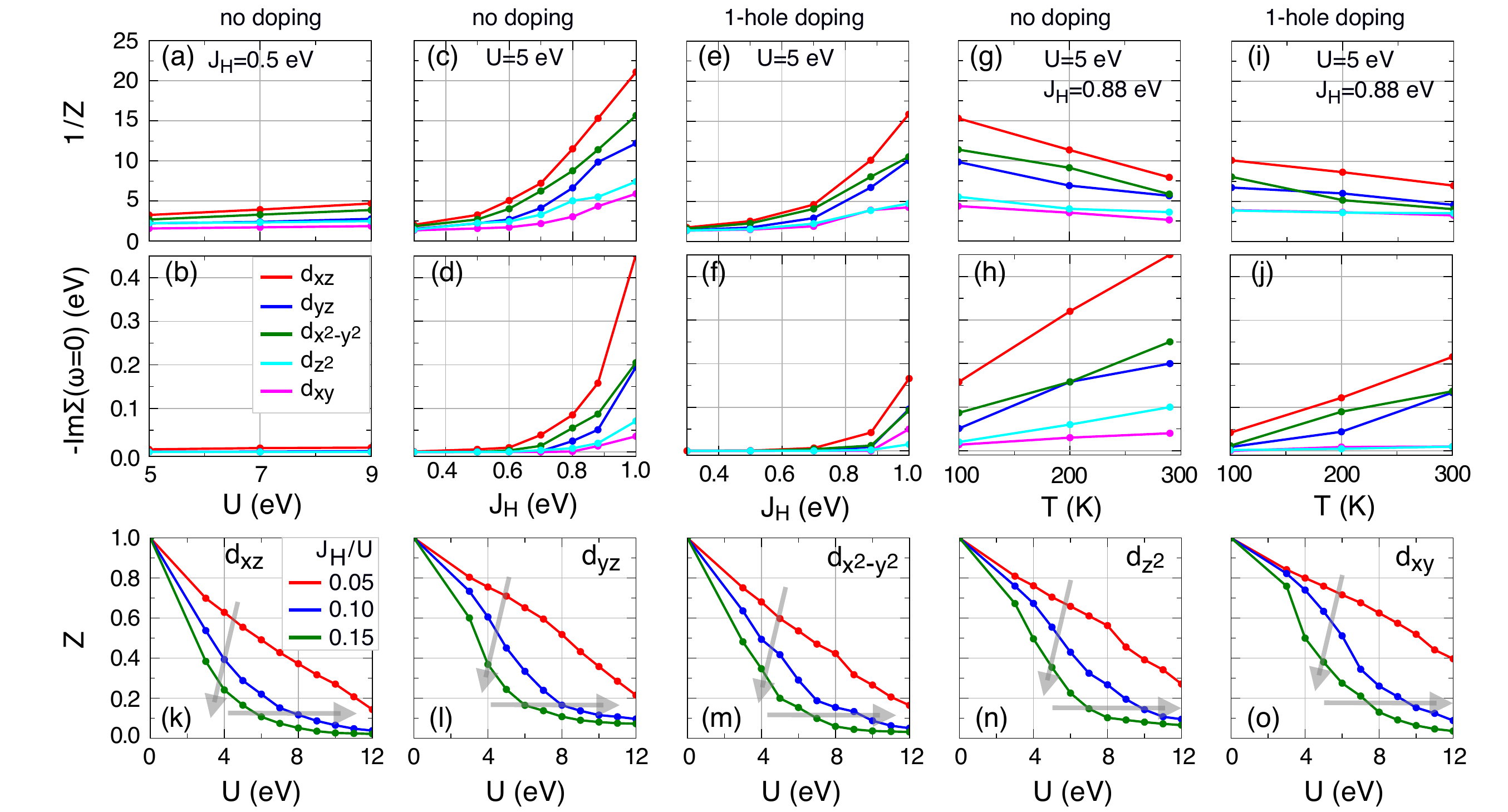}
    \caption{Heavy-fermions in CsCr$_3$Sb$_5$ revealed by DFT+DMFT calculations. Quasi-particle mass-enhancement $1/Z$ and effective scattering rate $-\text{Im}\Sigma(\omega=0)$ as functions of (a),(b) Hubbard $U$ at $J_H=0.5$ eV, $T=100$ K, (c)-(f) Hund's coupling $J_H$ at $U=5$ eV, $T=100$ K, (g)-(j) and temperatures at $U=5$ eV, $J_H=0.88$ eV. (k)-(o) Quasi-particle weight $Z$ as functions of $U$ at $J_H/U=$0.05, 0.1, 0.15, respectively, and $T=290$ K.}
    \label{fig:CCS}
\end{figure*}

\section{Results} 
The DFT calculated band structure and density of states (DOS) of CsCr$_3$Sb$_5$ are shown in Figs.~\ref{fig:band}(a) and (b). The fat-bands with Cr $d$-orbital characters are shown by colors in terms of the point-group of Cr-site ($D_{2h}$). We indeed find portions of incipient flat-bands with $d_{xz}$ ($B_{2g}$, red), $d_{yz}$ ($B_{3g}$, blue), $d_{x^2-y^2}$ ($A_{g}$, green) characters, which are 0.1$\sim$0.3 eV above $E_{\text{F}}$. These three orbitals also contribute dispersive bands around $E_{\text{F}}$. The $d_{z^2}$ ($A_{g}$, cryan) and $d_{xy}$ ($B_{1g}$, magenta) bands are very dispersive and contribute negligible DOS at $E_{\text{F}}$. There are also very dispersive Sb $p$-bands crossing $E_{\text{F}}$ (not shown as fat-bands). It should be emphasized that the total DOS at $E_{\text{F}}$ is, however, not large because those flat-band portions are away from $E_{\text{F}}$.

\begin{table}
    \centering
    \caption{Occupancy of Cr $d$-orbitals of CsCr$_3$Sb$_5$ calculated by DFT and DFT+DMFT  at $U$=5, $J_H$=0.88 eV, $T$=100 K.}
    \begin{ruledtabular}
    \begin{tabular}{c|cccccc}

                     & $d_{xz}$  & $d_{yz}$ & $d_{x^2-y^2}$ & $d_{z^2}$ & $d_{xy}$ & total \\
         \hline
         DFT, no-doping & 0.967 & 0.826 & 0.767 & 0.899 & 1.037 & 4.496 \\
         DFT, one-hole & 0.904 & 0.840 & 0.728 & 0.917 & 1.060 & 4.449 \\
         DMFT, no-doping & 0.908 & 0.837 & 0.845 & 0.831 & 0.868 & 4.289 \\
         DMFT, one-hole & 0.890 & 0.827 & 0.811 & 0.828 & 0.872 & 4.228 \\
    \end{tabular}
\end{ruledtabular}
    \label{tab:occu}
\end{table}

As shown in Table.~\ref{tab:occu}, our DFT and DFT+DMFT calculations yield about 4.5, 4.3 $d$-electrons per Cr-site, respectively, which are almost equally distributed among the five $d$-orbitals. This is nominally one-electron less than the half-filling, characteristic of Hund's metal. The quasi-particle mass-enhancement $m^*/m^{\text{DFT}}=1/Z$ ($Z$ is quasi-particle weight) and effective scattering rate $-\text{Im}\Sigma(\omega=0)$ calculated by DFT+DMFT are shown in Fig.~\ref{fig:CCS}. At moderate $J_\text{H}$ (0.5 eV), the largest mass-enhancement is only 2.5 $\sim$ 5 at relatively large Hubbard $U$ (5 $\sim$ 9 eV) comparing to the narrow bandwidths of those incipient flat-bands ($<$ 0.5 eV) [Fig.~\ref{fig:CCS}(a)]. The scattering rates are almost zero [Fig.~\ref{fig:CCS}(b)]. These are not consistent with the bad-metal and heavy-fermion behaviors of CsCr$_3$Sb$_5$~\cite{Caoguanghan:2023}. Therefore, the Hubbard $U$ cannot induce very strong correlations even with the presence of those incipient flat-bands, in contrast to what we usually expect. As shown in Fig.~\ref{fig:CCS}(c), the correlations are very sensitive to Hund's coupling $J_\text{H}$ and show very strong orbital differentiation at large $J_\text{H}$. The three orbitals $d_{xz}$, $d_{yz}$ and $d_{x^2-y^2}$ that contribute the incipient flat-bands are much more correlated than the $d_{z^2}$ and $d_{xy}$ orbitals. At $U$=5 eV and $T$=100 K, very large mass-enhancement (11 $\sim$ 21) is induced for the $d_{xz}$ orbital at typical $J_\text{H}$ (0.8 $\sim$ 1.0 eV) for $3d$-electron materials~\cite{yin:2011}. This indicates moderately heavy-fermions in CsCr$_3$Sb$_5$, consistent with the large $\gamma$ observed by experiment~\cite{Caoguanghan:2023}. Meanwhile, very large scattering rate is also induced [Fig.~\ref{fig:CCS}(d)], which explains the bad-metal behavior~\cite{Caoguanghan:2023}. On the contrary, the mass-enhancement of $d_{z^2}$ and $d_{xy}$ orbitals are only 3$\sim$8 at $J_\text{H}=$0.8$\sim$1.0 eV. 

In the pioneering work by Medici \textit{et al.}, it shows that $J_\text{H}$ exhibits ``Janus-Faced'' influence in electron correlations in Hund's metal~\cite{medici:2011}. On one hand, quasi-particle weight $Z$ is suppressed with increasing $J_\text{H}$. On the other hand, the critical $U$ for Mott transition is strongly increased at larger $J_\text{H}$. As a result, $Z$ displays a long tail as function of $U$, i.e., it shows strongly correlated bad metal behavior in a wide range of $U$ (see Fig.1(b) in~\cite{medici:2011}). For CsCr$_3$Sb$_5$, we show $Z$ as functions of $U$ at $J_\text{H}/U$=0.05, 0.1, 0.15, in Figs.~\ref{fig:CCS}(k)-(o). As indicated by the grey arrows, it indeed shows ``Janus-Faced'' behavior. We should note that, different from the degenerate 3-orbital model~\cite{medici:2011}, the five $d$-orbitals in CsCr$_3$Sb$_5$ are split due to crystal field, so Mott transition is much harder in CsCr$_3$Sb$_5$, and it does not occur at reasonable $U$ (up to 12 eV) for $d$-electron material. However, the overall ``Janus-Faced'' behavior is similar to that found in Ref.~\cite{medici:2011}. In particular, $Z$ displays a long tail in a wide range of $U$. Therefore, these results establish CsCr$_3$Sb$_5$ as a typical Hund's metal, and indicate the important roles of those incipient flat-band portions in enhancing the orbital differentiation via Hund's coupling $J_\text{H}$.

\begin{figure*}
    \centering
    \includegraphics[width=1.0\textwidth]{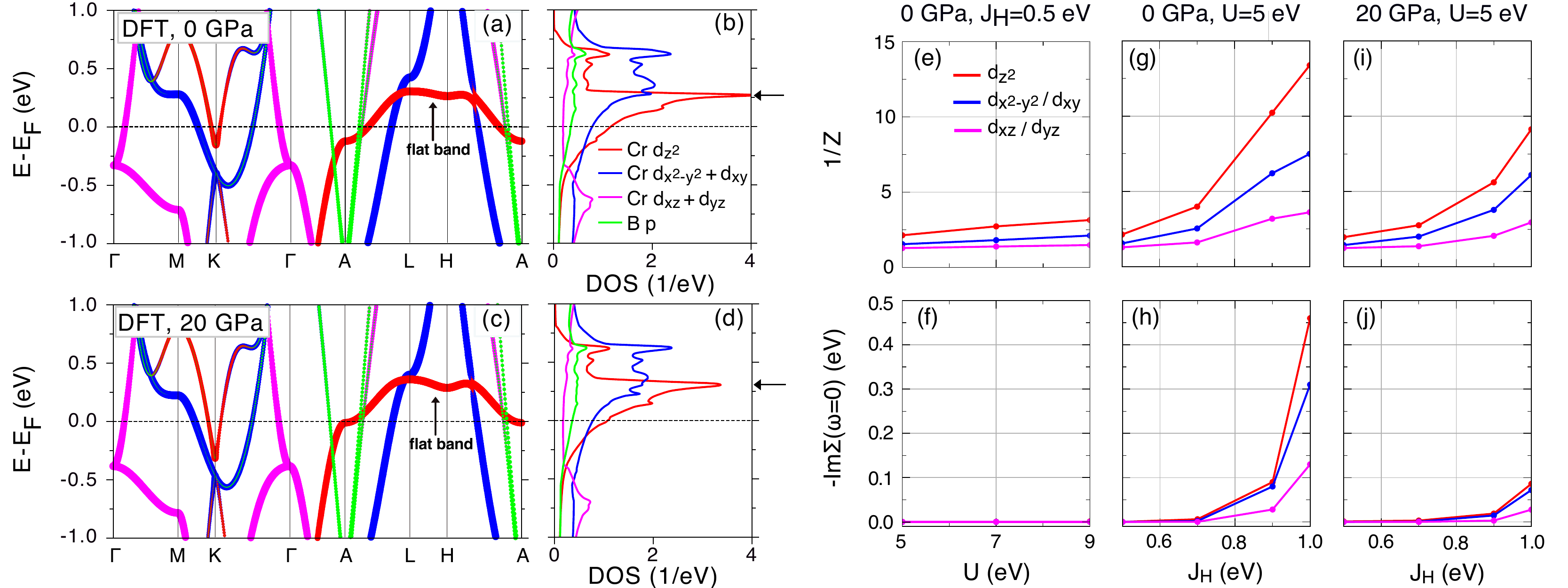}
    \caption{Incipient flat-band and heavy-fermions in CrB$_2$. (a)-(d) DFT calculated band structures and DOS. DFT+DMFT calculated mass-enhancement and scattering rate as functions of (e),(f) Hubbard $U$, (g)-(j) Hund's coupling $J_H$, at $T=100$ K.}
    \label{fig:CB}
\end{figure*}

\begin{figure*}
    \centering
    \includegraphics[width=1.0\textwidth]{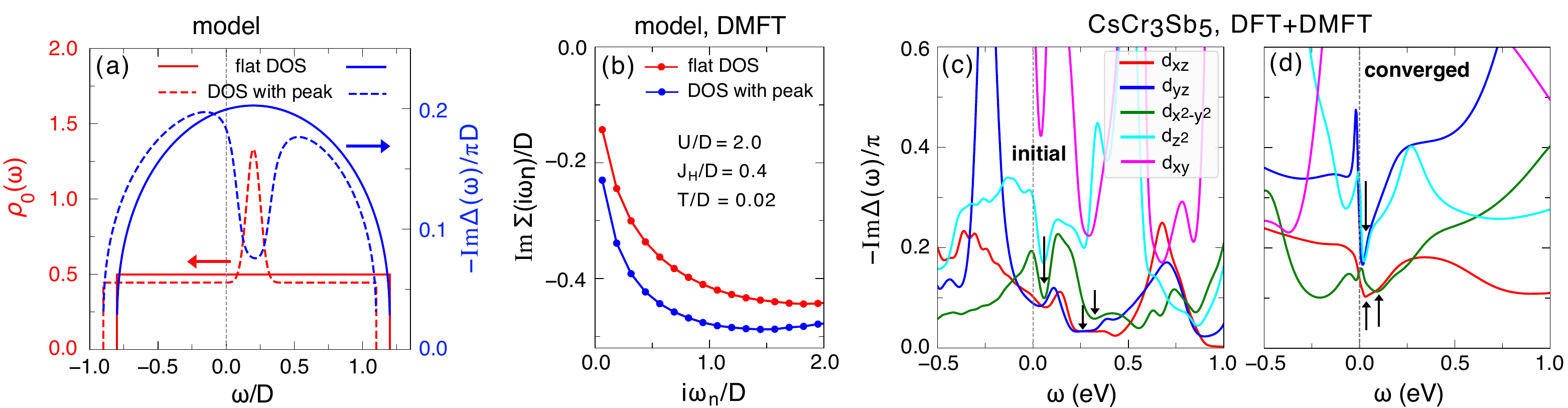}
    \caption{(a) Non-interacting DOS $\rho_0(\omega)$ (red) and hybridization strength $-\text{Im}\Delta(\omega)/\pi$ (blue) of a degenerate 5-orbital model occupied by 4 electrons, for flat DOS (solid) and DOS with a peak (dahsed), respectively. $D$ is half-bandwidth. (b) DMFT calculated self-energy $\text{Im}\Sigma(i\omega_n)$ at Matsubara frequency for this model. (c),(d) The initial and converged hybridization functions $-\text{Im}\Delta(\omega)/\pi$ of CsCr$_3$Sb$_5$ in DFT+DMFT calculation. The black arrows indicate dips in hybridization functions, which correspond to peaks in the DOS as shown in Figs.~\ref{fig:band}(b),(f).
     }
    \label{fig:hyb}
\end{figure*}

To further reveal the roles of the incipient bands, we apply one-hole doping to modify their positions, which are shown in Figs.~\ref{fig:band}(c),(d). The flat-bands become a little further away from $E_{\text{F}}$. As a result, the correlations are significantly reduced [Figs.~\ref{fig:CCS}(e),(f)]. For example, the mass-enhancement and scattering rate of $d_{xz}$ orbital decrease by about 33\% (from 15 to 10) and 73\% (from 0.15 to 0.04 eV), respectively, at $U$=5, $J_\text{H}$=0.88 eV and $T=$100 K. The orbital differentiation also becomes weaker. We note that, as shown in Table~\ref{tab:occu}, the total occupancy of Cr $d$-orbitals decreases only by 0.05$\sim$0.06 electrons per Cr-site under one-hole doping, indicating most of the doped holes ($\sim$80\%) enter Sb $p$-orbitals. Such small change in $d$-orbital occupancy can not account for such dramatic change of the correlations. Therefore, the large reduction of the correlations can only be attributed to the changes of the positions of those incipient flat-band portions. This indicates that the electron correlations and thus the magnetism are very sensitive to those incipient flat-band portions, so they should be one of the key factors responsible for the pressure-induced superconductivity in CsCr$_3$Sb$_5$. In reality, the one-hole doping can be realized by replacing one Sb-site with Ge (see Fig. S1 and Table S1.~\cite{suppl}).

The DFT calculated band structures and DOS of CrB$_2$ are shown in Figs.~\ref{fig:CB}(a)-(d). The Cr $d$-bands are split into three groups in terms of the point-group of Cr site ($D_{6h}$): $d_{z^2}$ ($A_{1g}$, red), $d_{x^2-y^2/xy}$ ($E_{2g}$, blue) and $d_{xz/yz}$ ($E_{1g}$, magenta). There are also very dispersive B $p$-bands (green) crossing $E_{\text{F}}$. At the BZ boundary ($k_z=\pi$), there is an incipient flat-band with $d_{z^2}$ character, which is about 0.25 eV above $E_{\text{F}}$ [Figs.~\ref{fig:CB}(a),(b)]. 

Similar to CsCr$_3$Sb$_5$, our DFT+DMFT calculations yield about 4.2 $d$-electrons per Cr-site for CrB$_2$ (see Table S3~\cite{suppl}).  At moderate $J_\text{H}$ (0.5 eV), the largest mass-enhancement ($d_{z^2}$ orbital) is only 2.1 $\sim$ 3.1 at $U$=5 $\sim$ 9 eV [Fig.~\ref{fig:CB}(e)], and the scattering rate is almost zero [Fig.~\ref{fig:CB}(f)]. While, the electron correlations are very sensitive to $J_{\text{H}}$ and show strong orbital differentiation [Figs.~\ref{fig:CB}(g),(h)]. The $d_{z^2}$ orbital that contributes the incipient flat-band is the most correlated one. For example, it has very large mass-enhancement ($\sim$10) and scattering rate at $U$=5 eV, $J_{\text{H}}$=0.9 eV, $T$=100 K, which are consistent with the bad-metal and heavy-fermion behaviors observed by the experiment~\cite{Bauer:2014}. It also shows ``Janus-Faced'' effect (see Fig. S5 in~\cite{suppl}). Therefore, CrB$_2$ is also a typical Hund's metal. At 20 GPa, the most pronounced change is that the incipient flat-band becomes slightly broader and further away from $E_{\text{F}}$ [Figs.~\ref{fig:CB}(c), (d)], and as a result, the correlations are significantly reduced. For example, the mass-enhancement of $d_{z^2}$ orbital decreases by 45\% (from 10.2 to 5.6) [Figs.~\ref{fig:CB}(i)]. The orbital differentiation also becomes weaker. This again indicates the important roles of the incipient flat-bands in driving the strong electron correlations and heavy-fermions. 

We can establish a general connection of incipient flat-band portions to the electron correlations via the hybridization function $\Delta(\omega)$. In the non-interacting limit, the hybridization strength is inversely proportional to DOS, $-\text{Im}\Delta(\omega)/\pi\propto 1/\rho_0(\omega)$~\cite{Mravlje:2011}. Hence, large peaks in DOS near $E_F$ lead to dips in $-\text{Im}\Delta(\omega)/\pi$. Smaller hybridization will result in stronger correlation in metals. This is demonstrated by a degenerate 5-orbital model occupied by 4 electrons [Fig.~\ref{fig:hyb}(a)]. Comparing to a completely flat DOS (solid red), the presence of a large peak above $E_F$ (dashed red) leads to a large dip in $-\text{Im}\Delta(\omega)/\pi$ (dashed blue). As a result, the correlation indeed becomes much stronger, as shown by the imaginary part of self-energy at Matsubara frequency calculated by DMFT [see Fig.~\ref{fig:hyb}(b)]. We plot $-\text{Im}\Delta(\omega)/\pi$ of CsCr$_3$Sb$_5$ in Figs.~\ref{fig:hyb}(c), (d). There are dips around 0.06, 0.25, 0.3 eV above $E_F$ in the initial hybridization function, which correspond to the flat-band-induced peaks in the non-interacting DOS [see Fig.~\ref{fig:band}(b)]. The DMFT self-consistence further pushes these dips much closer to $E_F$ in the converged hybridization function [Fig.~\ref{fig:hyb}(d)].

The key physics in Hund's metal is the screening of large local moments via the hybridization function. Smaller hybridization strength (dips) near $E_F$ causes the screening less effective and results in bad metal behavior at high temperature, and a full screening can only be achieved at very low temperature. This leads to the emergence of very heavy renormalized electrons at low temperature, in the same spirit of Kondo physics. In other words, the presence of incipient flat-band portions (sharp peaks in DOS) in Hund's metal makes the Hund's metal physics much closer to Kondo physics. This is the key finding of our work. This is supported by the Kondo-like crossover shown in Figs.~\ref{fig:CCS}(g) and (h). With decreasing temperatures, the scattering rates significantly decrease [Fig.~\ref{fig:CCS}(h)] and quasi-particles emerge with dramatic enhancement of their mass [Fig.~\ref{fig:CCS}(g)]. While, the Kondo-like effect becomes weaker when the flat-bands are pushed further away from $E_{\text{F}}$ by the one-hole doping [Figs.~\ref{fig:CCS}(i),(j)].  

Therefore, as long as the non-interacting DOS have sharp peaks near $E_F$ in Hund's metal, it will significantly enhance the electron correlations. Such peaks are not required to be exactly at $E_F$. This provides large flexibility to induce such physics in real materials. One rich platform is the frustrated lattices with widespread incipient flat-band portions. Since the key physics is local, the details of band structures that lead to sharp peaks is less important, and the single-site DFT+DMFT has captured the main physics. 

\section{Conclusion and Discussion} 
To summarize, we show that the presence of incipient flat-band portions in Hund's metals will induce dips in hybridization function, which further enhance the Kondo-like effect of Hund's metals, resulting in moderately $d$-electron heavy-fermions. Since it does not require that the flat-bands are entirely flat in the whole BZ and exactly sitting at $E_{\text{F}}$, the frustrated Hund's metals with incipient flat-bands provide a flexible route for generating $d$-electron heavy-fermions and a good environment for the emergence of unconventional superconductivity. A large family of materials with kagome, triangular, pyrochlore lattices consisting of Cr (d$^4$), Ru (d$^4$), Fe (d$^6$) ions are promising candidates. 

Our results naturally explain the origin of the bad-metal and heavy-fermions behaviors in the normal states of the superconducting kagome CsCr$_3$Sb$_5$ and triangular CrB$_2$~\cite{Caoguanghan:2023,Bauer:2014}, which pay the way for understanding their unconventional superconductivity mechanism. Previous studies suggest that incipient flat-bands are helpful for enhancing electron pairing~\cite{Matsumoto:2018,Aoki:2020,Kazuhiko:2017}, so they may be also crucial for the superconductivity in CsCr$_3$Sb$_5$ and CrB$_2$.

We note that there is also incipient flat-band portion with $d_{z^2}$ character in the recently studied nickelate superconductors, La$_3$Ni$_2$O$_7$ and La$_4$Ni$_3$O$_{10}$. DMFT calculations also find signature of Hund's metal behaviors~\cite{Ouyang:2024,Jingxuan:2024}.  ARPES experiment~\cite{ZhouXinJiang:2024} finds a very large mass-enhancement (5$\sim$8) for the $d_{z^2}$ orbital, indicating heavy $d$-electrons. This can also be explained by our theory. Ref.~\cite{YiHengTian:2024} proposed that Hund's correlation is very important for the superconductivity in La$_3$Ni$_2$O$_7$, and it should be also crucial for CsCr$_3$Sb$_5$ and CrB$_2$.


\section{Acknowledgement} This project was supported by the National Natural Science Foundation of China (No. 12174365), the Innovation Program for Quantum Science and Technology (No. 2021ZD0302800) and the New Cornerstone Science Foundation. The calculations were performed in Hefei Advanced Computing Center, China.

\textit{Note added.} Recently, we learned that another paradigm for finding $d$-electron heavy-fermions based on Hund's metal has been proposed by Crispino \textit{et. al.}~\cite{Crispino:2023}, by doping towards half-filling of a Hund's metal to enhance the orbital differentiation and reach extremely heavy mass-enhancements with a given orbital character.

\bibliography{main}

\end{document}